\documentclass[conference]{IEEEtran}

\usepackage{amsmath,amssymb,amsfonts}
\usepackage{graphicx} \graphicspath{ {Figures/} }
\usepackage{tabularx}
\usepackage{url}
\usepackage{xcolor}
\usepackage{algorithm,algpseudocode}
\usepackage{nicefrac}
\usepackage{lipsum}
\usepackage{cite}
\usepackage{multirow}

\begin{document}

\title{Bitcoin Double-Spending Attack Detection using Graph Neural Network}

\author{
    \IEEEauthorblockN{
    Changhoon Kang\IEEEauthorrefmark{1},
    Jongsoo Woo\IEEEauthorrefmark{2},
    and James Won-Ki Hong\IEEEauthorrefmark{1}
}
\IEEEauthorblockA{
    \IEEEauthorrefmark{1}Department of Computer Science and Engineering, POSTECH, South Korea
}
\IEEEauthorblockA{
    \IEEEauthorrefmark{2}Graduate School of Information Technology, POSTECH, South Korea
}
}

\IEEEoverridecommandlockouts

\IEEEpubid{\makebox[\columnwidth]{978-8-3503-1019-1/23/\$31.00~\copyright2023 IEEE \hfill} \hspace{\columnsep}\makebox[\columnwidth]{ }}

\maketitle

\IEEEpubidadjcol

\begin{abstract}
Bitcoin transactions include unspent transaction outputs (UTXOs) as their inputs and generate one or more newly owned UTXOs at specified addresses. Each UTXO can only be used as an input in a transaction once, and using it in two or more different transactions is referred to as a double-spending attack. Ultimately, due to the characteristics of the Bitcoin protocol, double-spending is impossible. However, problems may arise when a transaction is considered final even though its finality has not been fully guaranteed in order to achieve fast payment. In this paper, we propose an approach to detecting Bitcoin double-spending attacks using a graph neural network (GNN). This model predicts whether all nodes in the network contain a given payment transaction in their own memory pool (mempool) using information only obtained from some observer nodes in the network. Our experiment shows that the proposed model can detect double-spending with an accuracy of at least 0.95 when more than about 1\% of the entire nodes in the network are observer nodes.

\end{abstract}


\section{Introduction}
\label{sec:intro}
Bitcoin~\cite{nakamoto2008bitcoin} transactions use the concept of unspent transaction output (UTXO). UTXO is the Bitcoin balance received by transactions included in the blockchain in the past. An owner of this UTXO can send bitcoins by using it as an input when creating a new transaction. At this time, attempting to use the same UTXO more than once in multiple transactions is called a double-spending attack~\cite{karame2015misbehavior}. For convenience, in this paper, we refer to a transaction generated for fast payment and known to the merchant as $\mathbf{tx}_{pay}$, and a transaction to double-spend UTXOs that have already been used as inputs of $\mathbf{tx}_{pay}$ as $\mathbf{tx}_{attack}$. Eventually, double-spending is impossible in Bitcoin because only one chain with the longest block extension is recognized as valid according to Bitcoin's longest chain rule. For fast payment, if the merchant approves $\mathbf{tx}_{pay}$ without sufficient confirmation, and the attacker creates $\mathbf{tx}_{attack}$, the merchant may not be paid.

In this paper, we propose an approach to detect Bitcoin double-spending attacks using a graph neural network (GNN)~\cite{scarselli2008graph}. The Bitcoin P2P network can be considered as an undirected graph where peer nodes are vertices and their connections are edges. We use GNN to predict whether there is a $\mathbf{tx}_{attack}$ across the network for $\mathbf{tx}_{pay}$, using only information obtained from a few observer nodes that we can find in their mempools all unconfirmed transactions. We assume that double-spending does not occur if all nodes have $\mathbf{tx}_{pay}$. To the best of our knowledge, this work is a novel approach that uses GNN to predict the propagation status of an unconfirmed transaction in the Bitcoin P2P network.

\section{Related Work}
Many attempts are being made to achieve fast Bitcoin payments while preventing double-spending attacks. For instance, the Lightning network~\cite{poon2016bitcoin} uses a Layer-2 payment protocol to improve payment speed. Additionally, many other studies have proposed methods that require partial modification of the Bitcoin protocol~\cite{herrmann2012implementation,karame2012two, karame2012double,perez2019double,johnson2001elliptic}. However, it would be difficult to modify the Bitcoin protocol solely for this purpose. Furthermore, most attempts are difficult for ordinary people who do not know much about blockchain, such as merchants. They would not be able to operate a payment channel for the Lightning network or manage peer connections of their own Bitcoin nodes to remain connected to arbitrary samples of nodes to make it difficult for attackers to successfully execute double-spending. In contrast, a platform called GAP600~\cite{gap600} provides real-time guarantee service for fast payment through risk scoring for each unconfirmed transaction through network monitoring. However, specific risk analysis and evaluation methods are not disclosed at all.


\section{Method}
\label{sec:method}
In this paper, we construct virtual Bitcoin networks to generate data through transaction propagation simulations. Although a GNN model is learned through graph structures that are different from the real Bitcoin network topology, it is possible to apply the learned model to the real Bitcoin network thanks to the characteristic of GNN.

In this work, we do not use the topology of the actual Bitcoin network, but instead create virtual Bitcoin networks with a similar structure. The number of nodes in the virtual Bitcoin networks is fixed at 14,000 by referring to the website bitnodes.io~\cite{bitnodes}, which captures and displays reachable Bitcoin nodes in real-time. Additionally, according to \cite{essaid2020bitcoin}, the Bitcoin network is not a random graph and has some community structures. Therefore, we assume that the Bitcoin network is similar to a scale-free network~\cite{barabasi2003scale}. A characteristic of the scale-free network is that a node with a higher degree gets more new connections. We use the Barabási-Albert model~\cite{barabasi1999emergence} to create virtual Bitcoin networks.

\begin{table*}[t]
\caption{Experimental results for 150, 200, and 250 observer nodes}
\label{tab:results}
\renewcommand{\arraystretch}{1.5}
\resizebox{\textwidth}{!}{%
\begin{tabular}{|c|c|c|c|c|c|c|c|c|c|}
\hline
\textbf{Observer Nodes} & \multicolumn{3}{c|}{\textbf{150}}  & \multicolumn{3}{c|}{\textbf{200}} & \multicolumn{3}{c|}{\textbf{250}} \\ \hline
\textbf{GNN Layer} & \textbf{GCN} & \textbf{GraphSAGE} & \textbf{GAT} & \textbf{GCN} & \textbf{GraphSAGE} & \textbf{GAT} & \textbf{GCN} & \textbf{GraphSAGE} & \textbf{GAT} \\ \hline
\textbf{CV Avg. Accuracy} & 0.9500 & 0.9571 & 0.9571 & 0.9557 & 0.9729 & 0.9786 & 0.9671 & 0.9729 & 0.9971 \\ \hline
\textbf{Test Accuracy} & 0.9733 & 0.9500 & 0.9533 & 0.9633 & 0.9733 & 0.9733 & 0.9900 & 0.9900 & 0.9933 \\ \hline
\textbf{Precision} & 0.9524 & 0.9143 & 0.9195 & 0.9313 & 0.9490 & 0.9490 & 0.9809 & 0.9809 & 0.9872 \\ \hline
\textbf{Recall} & 1.0 & 1.0 & 1.0 & 1.0 & 1.0 & 1.0 & 1.0 & 1.0 & 1.0 \\ \hline
\textbf{F1-score} & 0.9756 & 0.9552 & 0.9581 & 0.9644 & 0.9739 & 0.9739 & 0.9904 & 0.9904 & 0.9935 \\ \hline
\end{tabular}%
}
\end{table*}

Our GNN task is graph classification, which predicts whether every node on the graph has a given transaction ($\mathbf{tx}_{pay}$) in its mempool. Given information about whether some nodes, which are observers on the graph, possess $\mathbf{tx}_{pay}$, we predict whether all the nodes on the graph have $\mathbf{tx}_{pay}$ in their mempool. Observer nodes are randomly selected for each generated data. Our model uses two GNN layers, and after each GNN layer, a ReLU activation function to provide non-linearity and a dropout to prevent overfitting are added. In the last part, for graph classification, the softmax value of each node embedding is aggregated by calculating the mean value, and it is passed through a fully connected layer to produce an output for classification. In our case, the input graph uses the Bitcoin network as it is. We also use cross-entropy loss to train the model and existing three popular GNN algorithms for GNN layers: GCN~\cite{kipf2016semi}, GraphSAGE~\cite{hamilton2017inductive}, and GAT~\cite{velivckovic2017graph}.

In addition, we design node features to include information around the nodes on the graph. The number of neighboring nodes of each label (Non-Observers: 0.5, Observers with $\mathbf{tx}_{pay}$: 1, Observers without $\mathbf{tx}_{pay}$: 0) is used as features, and we also added the number of each combination consisting of the neighboring node label and the label of a node one hop further away from it as features.

\section{Experiment}
\label{sec:experiment}
We conducted experiments on a total of six cases by varying the number of observer nodes. For each case, we generated 1,000 datasets where the number of observer nodes was 10, 50, 100, 150, 200, and 250 out of the total 14,000 nodes. We split each dataset into 700 for training and 300 for testing. We also evaluated the model using 5-fold cross-validation on the training dataset.

\subsection{Observer Nodes: 150, 200, and 250}
Table~\ref{tab:results} presents the average accuracy of cross-validation for each model when the number of observer nodes is 150, 200, and 250, as well as the accuracy, precision, recall, and F1-score values for the test set. All three types of GNN layers demonstrated similar performance under the same number of observer nodes, and the higher the number of observer nodes, the higher the accuracy. In all cases, the accuracy was at least 0.95, and it is interesting to note that the recall had a value of 1.0. In our experiment, recall indicates how accurately the model predicted that there was no double-spending attack when there was no actual attack. In other words, the models trained in this experiment accurately predicted that there was no attack for all cases where there was no double-spending attack. Precision refers to the ratio of cases where there is actually no attack among those predicted by the model in our experiment that there is no double-spending attack. It is an essential indicator of the efficiency of double-spending attack detection for Bitcoin fast payment because the higher the precision value, the higher the rate of predicting that there is no attack only when there is actually no double-spending attack. As expected, the precision had a larger value as the number of observer nodes increased, and values of at least about 0.91 or higher were obtained.

\subsection{Observer Nodes: 10, 50, and 100}
When the number of observer nodes is 10, 50, and 100, we could not train the model. We found that train loss does not decrease even after repeated learning steps. We concluded that this is because the number of nodes with information, i.e., observer nodes, is too small to predict the state of the entire graph consisting of 14,000 nodes. In this experiment, it was possible to detect a double-spending attack with high accuracy only when the mempool data of at least about 1\% or more of the nodes was known.

\section{Conclusion}
\label{sec:conclusion}
This paper proposes a GNN-based approach to detect Bitcoin double-spending attacks by predicting attempts to double-spend UTXOs in payment transactions. The model achieved an accuracy of at least 0.95 by monitoring 150 or more observer nodes' mempool in a 14,000-node P2P network. However, training was inadequate with less than 100 observer nodes.

\section*{Acknowledgements}
This work was supported by Coinone and Smart HealthCare Program(www.kipot.or.kr) funded by the Korean National Police Agency(KNPA, Korea) [Project Name: Development of an Intelligent Big Data Integrated Platform for Police Officers’ Personalized Healthcare / Project Number: 220222M01]

\bibliographystyle{IEEEtran}
\bibliography{main.bib}
\end{document}